# Discovery of U-Carbon: Metallic and Magnetic

Hong Fang[1], Michael Masaki[2], Anand B. Puthirath[3], Jaime M. Moya[4], Guanhui Gao[3], Emilia Morosan[4], Pulickel M. Ajayan[3], Joel Therrien[2]*, Puru Jena[1]*

[1] Physics Department, Virginia Commonwealth University, Richmond, VA 23284-2000, USA.

[2] Department of Electrical and Computer Engineering, University of Massachusetts, Lowell, MA, USA.

[3] Department of Materials Science and NanoEngineering, Rice University, Houston, TX, 77005, USA.

[4] Department of Physics and Astronomy, Rice University, Houston, TX 77005, USA.

*Correspondence to: Joel_Therrien@uml.edu (J.T.); pjena@vcu.edu (P.J.)

**Abstract:** We report the discovery of a pristine crystalline 3D carbon that is magnetic, electrically conductive and stable under ambient conditions. This carbon material, which has remained elusive for decades, is synthesized by using the chemical vapor deposition (CVD) technique with a particular organic molecular precursor 3,3-dimethyl-1-butene ($C_6H_{12}$). An exhaustive computational search of the potential energy surface reveals its unique $sp^2$-$sp^3$ hybrid bonding topology. Synergistic studies involving a large number of experimental techniques and multi-scale first-principles calculations reveal the origin of its novel properties due to the special arrangement of $sp^2$ carbon atoms in lattice. The discovery of this U-carbon, named such because of its *unusual* structure and properties, can open a new chapter in carbon science.

Carbon, the building block of life on Earth, is a unique element in the periodic table. Due to its flexible bonding characteristics categorized by $sp^3$, $sp^2$ and $sp^1$ hybridization of the *s* and *p*-orbitals, it forms over ten million compounds whose properties are intimately related to their structures. Among these, diamond and graphite, two of the best-known three-dimensional (3D) forms of carbon with $sp^3$ and $sp^2$ bonding, respectively, display strikingly different structure and properties. Since the discovery of $C_{60}$ fullerene (*1*) we have witnessed the emergence of several new multi-dimensional carbon allotropes (*2-5*) exhibiting a range of spectacular properties. For decades, there has been a constant search for metallic and magnetic 3D carbon material with little success. The electrical conductivity ($2\sim3\times10^5$ S/m) of graphite in its basal plane comes from the 2D graphene layer with zero density of states at the Fermi level. A similar case is found in the bundle of armchair nanotubes, where the conductivity is from the isolated 1D nanotube along the bundle axis (*6*). A cubic phase of carbon formed under 3 Tera-Pascal ($3\times10^{12}$ Pa) pressure was reported to be metallic, but loses its stability when the pressure is removed (*7*). None of these carbon materials is magnetic. Although carbon magnetism has been reported in amorphous carbon structures (*5,8*), other studies suggest that the defect-mediated magnetism in the non-crystalline carbon is paramagnetic or has weak magnetic ordering (*9-10*). While hydrocarbon (*11*) and functionalized graphene (*12-13*) can also become magnetic, no pure crystalline magnetic carbon has ever been found. An early experiment claiming the observation of magnetic carbon was later found to be contaminated with magnetic metal impurities (*14*).

It is expected that a 3D metallic and magnetic carbon would be metastable with energy higher than that of graphite and would belong to certain local minimum in the potential energy surface (PES). We hypothesized that metastable phases can be achieved by limiting the accessible region in the PES during synthesis and "forcing" the produced structure into prescribed states. One possible way to realize this is to use selected molecular precursors rather than individual atoms in the formation. Given that the existing bonding structure inside the chosen molecular precursor will encounter an energy cost for any bond breaking and rearrangement, it may be energetically preferable for the precursor to maintain certain original bonding features when forming the metastable structure, especially in rapid reactions. This, to some extent, would limit the accessible region in the PES and novel carbon phases could form (*15-16*).

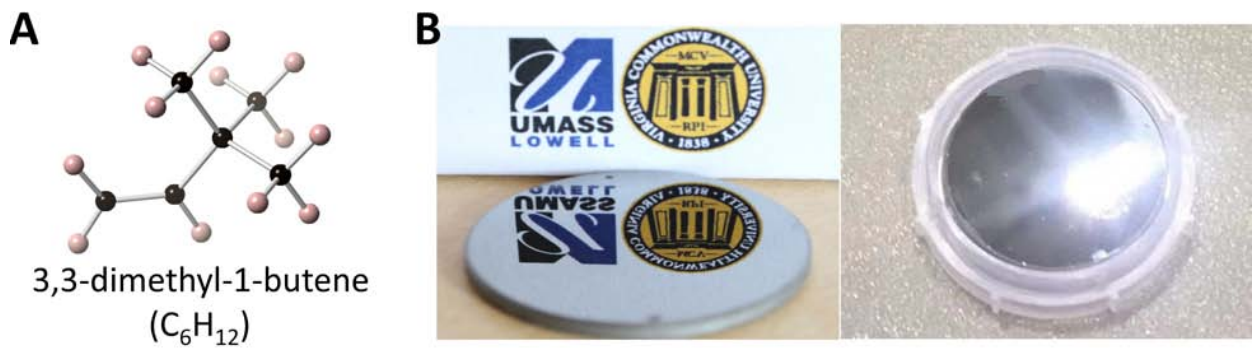

**Fig. 1.** Synthesis of U-carbon (UC). (**A**) Molecular structure of the precursor 3,3-dimethyl-1-butene ($C_6H_{12}$). (**B**) UC samples exhibit mirror-like appearance with metallic shine.

We chose 3,3-dimethyl-1-butene ($C_6H_{12}$) (Fig. 1A) as the molecular precursor with the aim of forming a new metastable $sp^2$-$sp^3$ hybrid bonding system. Samples were synthesized in a hot-wall CVD reactor operating at atmospheric pressure and temperatures ranging from 700-1000°C. The precursor 3,3-dimethyl-1-butene ($C_6H_{12}$) is added to the reactor via bubbling argon gas through it, upstream of the reactor. Injection of the precursor is started once the reactor has reached the desired temperature for growth (typically 800°C) and halted prior to cooling down. Growth was found to be substrate dependent, with metal oxides, copper, nickel, boron nitride and silicon oxide and nitride supporting growth. In contrast, un-oxidized silicon and glassy carbon do not show any indications of film growth (also see Section 1.1.1 in the Supplementary Information, SI). As a control, another form of hexane, cyclohexane with the identical chemical formula $C_6H_{12}$ but distinctively different molecular structure, was tested as precursor under the same growth conditions. From the measured FTIR and phonon Raman (Section 1.1.2 in SI), it is found that the resulting products from the two precursors are very different. This suggests that the precursor molecule's topology indeed plays a crucial role in the structure of the resulting carbon.

Thus, the synthesis supports a growth model where the hydrocarbon feedstock does not break down to atomic radicals and retains some of the original backbone structure. From the studies of thermal stability of hydrocarbons in the temperature range and flow rates comparable to the current settings of CVD growth, it is known that hexanes will not extensively break down (*17-18*). Another key distinction in the CVD here is the lack of hydrogen in the feedstock gas. Hydrocarbon cracking relies on hydrogen binding to the catalyst to lower the barrier energy for the hydrocarbon to undergo scission (*19-21*). The lack of large quantities of available hydrogen render such reaction pathway much less probable. With catalytic cracking suppressed, dehydrogenation becomes the dominant reaction at the catalyst surface (*22*), resulting in the formation of a carbon radical site on the molecule followed by bonding to the solid carbon.

The synthesized carbon material (Fig. 1B), named U-carbon (UC) because of its *unusual* structures and properties (to be discussed later), shows a high reflectivity from far UV to mid IR (Section 1.1.3 in SI). Unlike the appearance of amorphous carbon samples, the U-carbon samples are macroscopically uniform in nature and shine like metals with ultra-smooth surfaces (Fig. S3 in SI). There is no hydrogen left in the sample (Section 1.1.4 in SI), suggesting complete dehydrogenation of the precursor molecules during the synthesis. The measured XRD (Fig. S5 and Section 1.1.5 in SI) exhibits clear and distinctive peaks throughout the region, indicating a crystalline nature of the sample. The unusual broadness yet high intensity of the first peak around 26° suggests that it should be contributed by multiple, rather than a single, sets of crystal planes. This could correspond to a group of structures due to different stacking configurations or level of binding, leading to slightly different d-spacing. No single crystalline carbon materials known can match the measured XRD (e.g. graphite or diamond as shown in Fig. S6 in SI). Thus, UC clearly represents a new form of carbon.

We carried out a comprehensive structure search for UC and matched the simulated XRD with the experiment. One method applied is to optimize topologically assembled precursor units (Section 1.2.1 in SI), which resulted in a layered structure (Fig. 2A). Each layer, called U-graphene, has a buckled structure with three equally separated sublayers of carbon atoms (Fig. S8 in SI). Structures searched by an unbiased global optimization method based on *individual atoms* (Section 1.2.1 in SI) resulted in a pure $sp^3$ bonded structure (Fig. 2D).

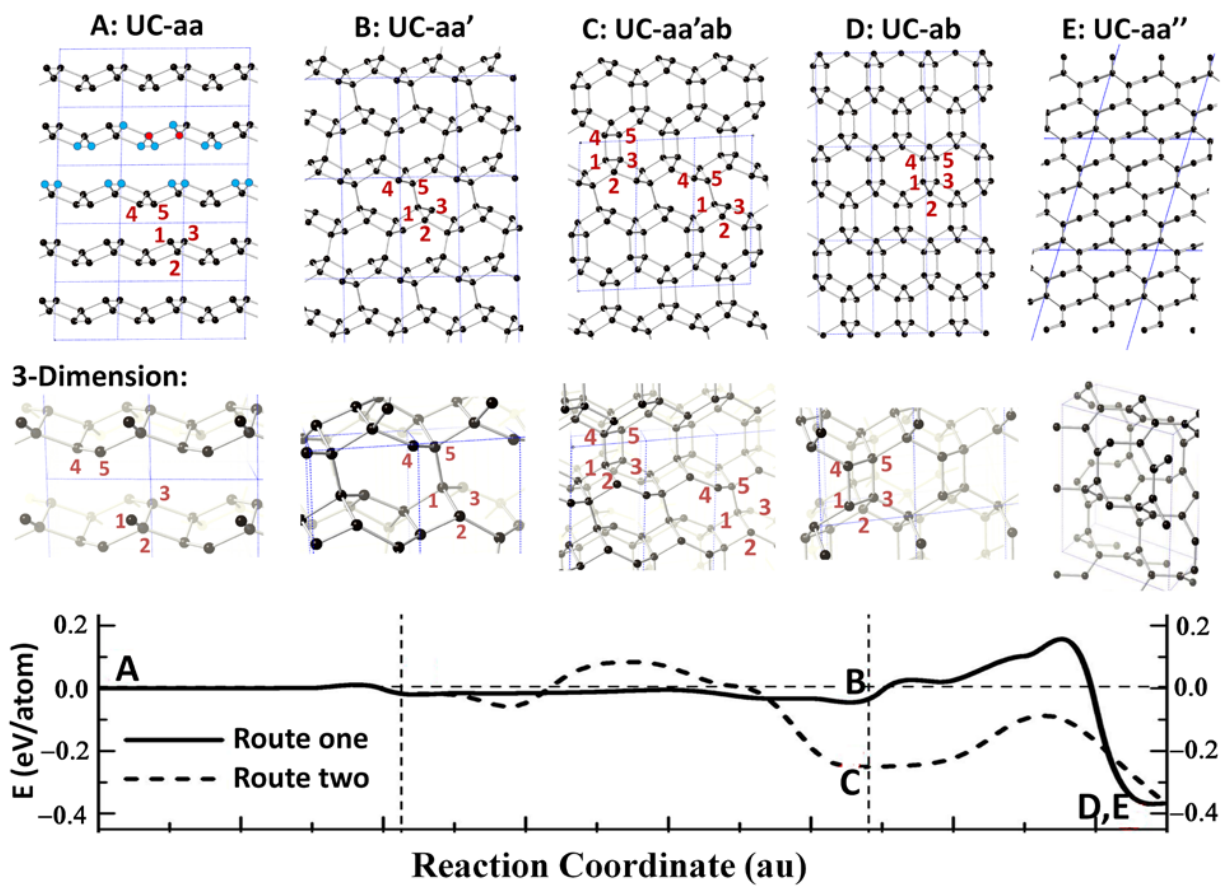

**Fig. 2.** Possible configurations of U-carbon (UC). These include UC-aa, UC-aa', UC-aa'ab, UC-ab and UC-aa". They originate from the possible stacking configurations of U-graphene (Fig. S2 in SI). In the UC-aa configuration, staggered-arranged $sp^2$ carbon atoms are shown in blue and the $sp^3$ carbon atoms in the unit cell are in red. The corresponding 3D views of the structures are shown with a tilted angle to illustrate the particular arrangement of the carbon atoms in the direction perpendicular to the paper. Note that the visually triangular arrangement of carbon (numbered as 1, 2 and 3) in the 2D view is actually composed by three carbon atoms in different planes parallel to the paper. This special geometrical arrangement is due to the uniquely buckled structure of U-graphene (Fig. S8 in SI), making the detected signals of the carbon atoms (1, 2 and 3 alike) overlapping with each other. Transitions between the structures can be realized by relative sliding and approaching of the neighboring layers, forming new bonds as demonstrated by the connections between the numbered atoms (4 and 5 connecting to 1 and 3). These configurations are shown along the calculated potential energy profiles of two identified reaction routes. The interconversion barriers separating UC-aa'ab, UC-ab and UC-aa" from the others are higher than 0.4 eV/atom.

We found that the layered and the $sp^3$ bonded structures actually belong to the same carbon material, caused by different stacking sequence of U-graphene. The AA-stacking of U-graphene results in the layered configuration (UC-aa, Fig. 2A), with $sp^2$ carbon atoms exposed outside to the neighboring layers. The AB-stacking of U-graphene results in the $sp^3$ bonded configuration (UC-ab, Fig. 2D). Other possible configurations include the $\frac{1}{2}$AB-stacking (UC-aa',
4

Fig. 2B) and the mixture of $\frac{1}{2}$AB- and AB-stacking (UC-aa'ab, Fig. 2C). Energetically, these configurations of UC correspond to a slew of metastable states ( which are lattice-dynamically stable as shown in Fig. S10 in SI) that are connected by continuous transitions (Movie S1 in SI), with the UC-aa and UC-ab as the starting and the end points, respectively (Fig. 2 and Fig. S9 in SI). Study on the kinetic stability of these competing UC configurations reveals that only UC-aa and UC-ab are stable at high temperatures (Section 1.2.2 and Fig. S11-12 in SI). UC-aa' undergoes a phase transition at 600 K (Fig. S13 in SI) to a pure $sp^2$ bonded structure, called UC-aa" (Fig. 2E), which becomes stable at high temperatures (Fig. S13 in SI). Therefore, it is possible to "purify" these three configurations using high temperatures.

Bearing this in mind, we annealed the sample at 1200°C under argon followed by cooling to room temperature. Major changes observed in the measured XRD of the annealed sample can be well explained by the simulated XRDs of the UC configurations (Fig. S5 in SI). The peaks around 35° are greatly weakened upon annealing, due to the disappearance of the UC-aa configuration with the (101) and (10$\bar{1}$) crystal planes. The peaks around 44° become prominent due to the prevalence of the UC-ab and UC-aa" configurations in the annealed sample. The calculated Raman/FTIR active modes based on the UC-ab and UC-aa" configurations also agree well with the measured ones (Section 1.2.2 in SI). Further analysis of the Raman spectra taken at two laser wavelengths suggests that the annealed UC sample has $sp^2$ and $sp^3$ bonds interspersed to a large degree and is not composed of phase segregated regions of graphitic or diamond-like materials. The degree of defects in the sample is found to be low (Section 1.1.6 in SI). The electron energy loss spectrum (EELS) (Fig. 3A) again shows that the sample is a $sp^2$–$sp^3$ hybrid system as expected from the UC configurations. Unlike the single broaden peak of amorphous carbon, the EELS of UC exhibits distinct peaks which are clearly different from those of graphite or diamond in terms of both position and relative height.

Next, the sample is investigated via high-resolution transmission electron microscopy (HRTEM) at different length scales (Fig. 3B and Section 1.1.7 in SI). The electron diffraction patterns clearly show the crystalline nature of UC (Fig. 3B). As shown in Fig. 3C, the measured XRD can be well described by the UC-ab, UC-aa" and UC-aa'ab configurations from the Rietveld analysis (Section 1.1.8 in SI). The measured lattice parameters are in good agreement with those calculated from first principles (Table S3 in SI). The broadened first peak between 23°~30° in the XRD is contributed by all three configurations of UC, including the crystal planes (110) and (101) of UC-aa", (110) of UC-ab, as well as (11$\bar{1}$), (11$\bar{2}$), (110), (20$\bar{2}$), (003) and (11$\bar{3}$) of UC-aa'ab. The two major peaks around 38° and 44° are from the UC-aa" and UC-ab configurations. Other relatively small yet distinctive peaks around 33°, 65° and 69° are all from the UC-aa'ab configuration. Absence of amorphous diffraction peaks and SAED patterns without diffused ring patterns further confirm that little amorphous phase exists in the sample (*23*).

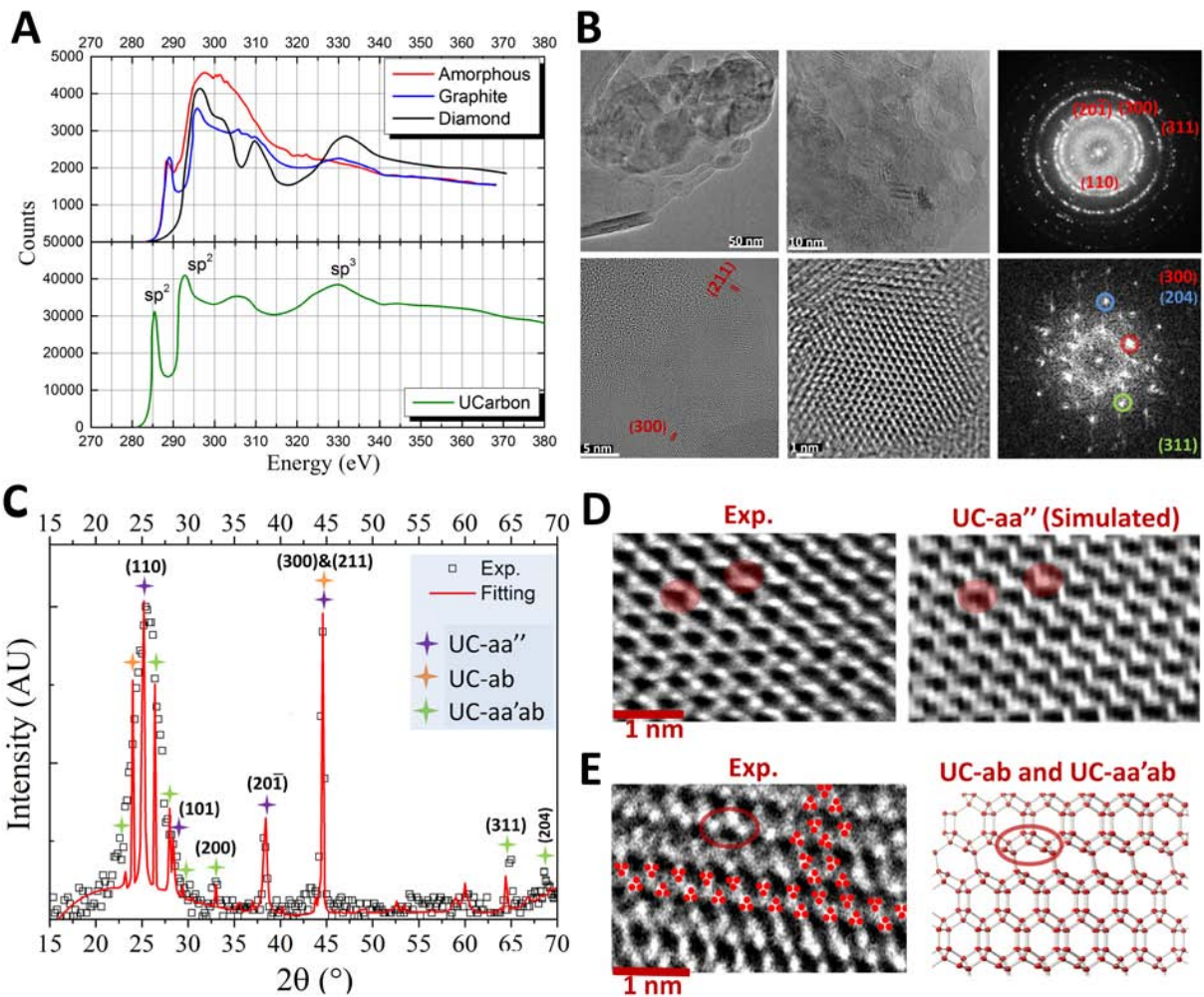

**Fig. 3.** (**A**) The electron energy loss spectrum (EELS) of the annealed UC sample, where the signatures of $sp^2$ and $sp^3$ bonded carbon are well evident. Presented for comparison are the EELS of amorphous carbon, graphite and diamond which are clearly different from that of UC. (**B**) High-resolution transmission electron microscopy (HRTEM) images (at different length scales) of UC and the electron diffraction patterns that clearly show the crystalline nature of UC. Each electron diffraction pattern corresponds to the TEM image on its left. The identified crystal planes and the corresponding d-spacing values are listed in Table S4 in SI. (**C**) Rietveld analysis on the measured XRD of the annealed sample using the configurations of UC-aa", UC-ab and UC-aa'ab. Miller indices are assigned accordingly to the XRD peaks. (**D**) Good match between the TEM image and the simulated one from the UC-aa" configuration (see Fig. S16 in SI for the corresponding atomic configuration), where the structure features highlighted in red are well reproduced. (**E**) Local structures from the measured TEM match with the UC-ab and the UC-aa'ab configurations. The detected signals are from the apparently overlapping carbon atoms when looking down the direction perpendicular to the paper (Fig. 2). The same structure feature in the TEM image and in the model structure is indicated by the red circle.

The UC-aa", UC-ab and UC-aa'ab configurations are identified in the HRTEM images. Fig. 3D shows good agreement between the experimental TEM image and the simulated one. Fig.

3E shows the TEM image that matches the UC-ab and UC-aa'ab configurations, characterized by fully or partially connected triangular-shaped atomic groups. According to the 3D views of the configurations of UC shown in Fig. 2, each triangular-shaped atomic group is actually composed by three columns of carbon atoms in different planes parallel to the paper. Such specific geometrical arrangement, due to the buckled configuration of U-graphene (Fig. S8 in SI), leads to overlapping of the detected signals of the carbon atoms in the same column but on the different planes.

Electrical transport properties of UC are studied by measuring the temperature-dependent resistivity (Section 1.1.9 in SI). At room temperature (RT), the sample can reach an electrical conductivity of $4.83(\pm0.02)\times10^5$ S/m which is about twice that of the graphite in its basal plane and is comparable to that of stainless steel. After annealing, the measured resistivity decreases on cooling from 400 to 375 K before increasing to the lowest measured temperature (Fig. 4A). The RT electrical conductivity of $4.10(\pm0.01)\times10^4$ S/m is about an order of magnitude smaller than that of the unannealed sample. This is expected due to the removal of the metallic UC-aa and UC-aa' configurations and simultaneous increase of the semiconducting UC-ab and UC-aa" configurations in the sample upon annealing (Fig. S17 in SI).

UC is found to be intrinsically magnetic under ambient conditions (Movies S10-S11 and Section 1.1.10 in SI). The magnetic susceptibility M/H of UC is measured on warming from 2 to 400 K, showing a splitting between zero field-cooled and field-cooled (H = 0.1T) data, which is characteristic of domain formations in ferromagnets (Fig. 4B). Magnetic isotherms measured both at 10 and 400 K saturate near 2.2 emu/g with small hysteresis (Fig. 4C and the inset). This suggests that UC is a soft ferromagnet with an ordering temperature above 400 K. To rule out any magnetic contaminant in the sample, ICP-OES tests with a sensitivity down to 10 parts per billion have been performed and the sample is confirmed to be pure without any trace of metal elements, such as Fe, Mn, Ni, Co *etc* (Table S4 in SI). The surface morphologies and elemental composition (EDAX) of the samples are also recorded, which shows that the sample is pure carbon with only a trace amount of oxygen coming from the atmosphere (Fig. S18 in SI). Applying a magnetic field of 0.01 T will decrease the resistivity of UC (Fig. 4A), leading to a negative magnetoresistance in the entire temperature range, which is characteristic of ferromagnets in the ordered state. The measured magnetoresistance isotherms at 10, 200, 300 and 400 K (Fig. 4D) further validate this. The measured Hall coefficient is positive in the entire temperature range (Fig. 4D), indicating dominant hole-like carriers with a carrier concentration of $3.17\times10^{20}/cm^3$ at RT.

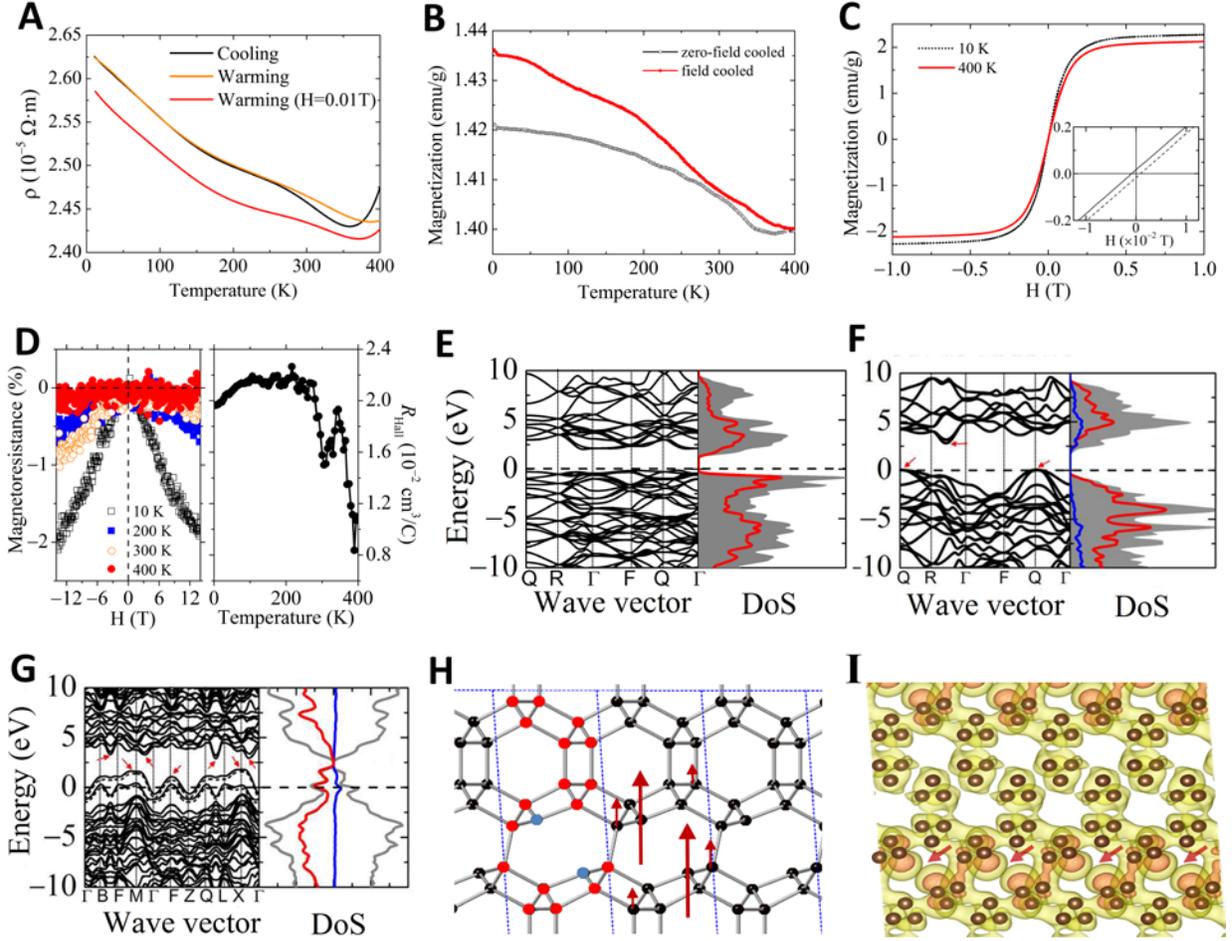

**Fig. 4.** Electrical transport and magnetic properties of the annealed UC sample. (**A**) Electrical resistivity of UC measured on cooling and warming as well as in a field of H = 0.01 T. The temperature dependence of the resistivity suggests that both holes and electrons are contributing to the conductivity. The reduced resistivity upon applying magnetic field is characteristic of a ferromagnet in the ordered state. (**B**) Magnetic susceptibility measured from 2 to 400 K in an applied field of 0.1 T. The splitting between the zero-field cooled data and the field cooled data with H = 0.1 T suggests domain formation in the material. (**C**) Magnetization isotherms measured at 10 and 400 K for fields swept from H = 0 to 1.125 T, 1.125 to −1.125 T, and −1.125 to 1.125 T. The saturated magnetization of the sample is about 2.2 emu/g. The inset shows a small hysteresis at 10 K. (**D**) Magnetoresistance measured at 10, 200, 300, and 400 K in field perpendicular to the current from −14 to 14 T. The negative magnetoresistance is typical for a ferromagnet in ordered state. The right panel shows the temperature-dependent Hall coefficient determined from temperature sweeps at H = ±14 T. (**E**) Calculated electronic band structure of UC-aa″ which is semiconducting. (**F**) Calculated electronic band structure of UC-ab which is semiconducting. (**G**) Calculated electronic band structure of UC-aa'ab which is metallic. The solid and dotted lines show the inequality of the two spins around the Fermi level (at zero energy). The red line in the density of states (DoS) shows that the dominant contribution to metallicity comes from the $p_z$ electrons which are also responsible for magnetism. The blue line shows the net spin around the Fermi level. (**H**) The ferromagnetic ground state of the UC-aa'ab configuration. The arrows show the projected magnetic moments on the carbon atoms. (**I**) Calculated charge density of the ferromagnetic ground state of UC-aa'ab. The arrows indicate the delocalized $p_z$ electrons.

The calculated electronic structures suggests that the UC-aa'ab configuration is responsible for the ferromagnetism of the sample, while the UC-aa" and UC-ab configurations are semiconducting and non-magnetic (Fig. 4E and F). UC-aa'ab is metallic and has a ferromagnetic ground state with a magnetic moment of 0.41 $\mu_B$ per unitcell (Fig. 4G). Its $p_z$ electrons from the $sp^2$ carbon atoms (Fig. 4H) are the origin of the metallicity and the dominant magnetic bearer (Fig. 4I). The origin of carbon magnetism is due to the specific topology of the local structure (Fig. 4H), where each $sp^2$ carbon is surrounded by three $sp^3$ carbon atoms, leading to a frustration of tetravalent bonding and an unpaired spin. Similar phenomenon is observed in the so-called magnetic carbon radical that can be stabilized by a "tetrapod" topology (*24*) present in schwarzites (*25*). It is worth noting that crystalline carbon structure having a calculated ferromagnetic ground state is extremely rare, with UC-aa'ab perhaps the only known case so far (*26-27*). According to the Rietveld analysis (Fig. 3C), the fractions of the different stacking configurations contained in the UC sample are 43%, 43% and 14% for UC-aa", UC-ab and UC-aa'ab, respectively. Thus, the theoretically estimated magnetization of the sample is about 3.0 emu/g (see Section 1.2.3 in SI) which agrees well with the experimental measurement of 2.2 emu/g (Fig. 4C). Compared to UC, the reported magnetization of amorphous carbon structures is much weaker at room temperature and will be further weakened upon graphitization (*28-29*).

In summary, we report the discovery of a 3D crystalline carbon material that is electrically conductive and ferromagnetic under ambient conditions. Synthesized using a particular molecular precursor, 3,3-dimethyl-1-butene, this *unusual* form of carbon, named U-carbon, originates from binding buckled $sp^2$-$sp^3$ hybridized carbon layers. Due to different stacking sequence, the U-carbon sample contains carbon configurations exhibiting semiconducting to metallic and ferromagnetic properties. Because of these unique properties, U-carbon is expected to have many scientific and technological applications. The use of selected molecular precursors that support a crystalline growth based on clustered rather than individual atoms is a paradigm shift in materials developments and can be used in the discovery of metastable materials beyond carbon.

**Acknowledgments:** We thank Dr. James Davenport of the Department of Energy for generous award of additional hours towards the year-end that made this study possible.

**Funding:** This work is partially supported by the U.S. Department of Energy, Office of Basic Energy Sciences, Division of Materials Sciences and Engineering under Award DE-FG02-96ER45579. Resources of the National Energy Research Scientific Computing Center supported by the Office of Science of the U.S. Department of Energy under Contract no. DE-AC02-05CH11231 is also acknowledged. A.B.P. thanks the Science & Engineering Research Board (SERB), India and Indo-US Science and Technology Forum (IUSSTF) for financial support in the form of postdoctoral fellowship. J.M.M. was supported by the National Science Foundation Graduate Research Fellowship under Grant DGE 18424893. J.M.M. and E.M. acknowledge support from NSF DMR 1903741.

**Author contributions:** H.F., J.T. and P.J. designed the research; H.F. and P.J. carried out the theoretical research and computational studies; M.M. and J.T. carried out the experimental synthesis; M.M., A.B.P., J.M.M., E.M., G.G., P.M.A. and J.T. carried out the experimental measurements; H.F., A.B.P., P.M.A., J.T. and P.J. analyzed the data; H.F., J.T. and P.J. wrote the paper. All authors contributed to editing the paper.